\documentclass[11pt]{article}

\usepackage[ps,dvips,all,frame,matrix]{xy}
\usepackage{amsmath}
\usepackage{amsthm}
\usepackage{epigraph}
\usepackage[a4paper]{geometry}
\usepackage{hyperref}

\usepackage{natbib}
\let\cite\citep
\newtheorem{theorem}{Theorem}
\newtheorem{axiom}[theorem]{Axiom}

\newtheorem{claim}[theorem]{Claim}

\newtheorem{definition}[theorem]{Definition}
\newtheorem{example}[theorem]{Example}

\newtheorem{proposition}[theorem]{Proposition}

\newtheorem*{mainthm}{Theorem of limited growth}
\newtheorem*{thermo}{Laws of Thermodynamics}
\newtheorem*{corzero}{Corollary of zero growth}

\usepackage{amssymb}

\begin{document}

\title{The Entropy Law and the impossibility \\of perpetual economic growth}
%\shortTitle{Entropy and the impossibility of perpetual growth}
\author{Ademar R. Romeiro\thanks{Unicamp - Institute of Economics, % R. Pit\'agoras 353, 13083-857
%\newline
\indent\hspace{0.5cm}
 \texttt{arromeiro@gmail.com}} 
\and 
Henrique N. S\'a Earp\thanks{Unicamp - Institute of Mathematics, %R. S\'ergio Buarque de Holanda 651, 13083-859
%\newline
\indent \hspace{0.35cm}\texttt{henrique.saearp@ime.unicamp.br}; 
%\newline\indent \quad both at: Cidade Universit\'aria, Campinas-SP, Brazil. %\newline
}}
\date{\today}
\maketitle
\vspace{-0.5cm}

\begin{abstract}
Every production-recycling iteration accumulates an inevitable proportion of its matter-energy in the environment, lest the production process itself would be a system in perpetual motion, violating the second law of Thermodynamics. Such high-entropy matter depletes finite stocks of ecosystem services provided by the ecosphere, hence are incompatible with the long-term growth in the material scale of the economic process. Moreover, the complex natural systems governing such stocks respond to depletion by possibly sudden environmental transitions, thus hindering markets' very ability to adapt to the new equilibrium conditions.
Consequently, uncertainty of critical resilience thresholds constrains  material economic growth.
\end{abstract}

\tableofcontents
\newpage

\setlength\epigraphwidth{9cm}
\renewcommand{\textflush}{flushepinormal}
\renewcommand{\epigraphsize}{\footnotesize}
\epigraph{\vspace{1cm}
The law that entropy always increases holds, I think, the supreme position among the laws of Nature. If someone points out to you that your pet theory of the universe is in disagreement with Maxwell's equations - then so much the worse for Maxwell's equations. If it is found to be contradicted by observation - well, these experimentalists do bungle things sometimes. But if your theory is found to be against the second law of thermodynamics I can give you no hope; there is nothing for it but to collapse in deepest humiliation.}
{Sir Arthur S. Eddington (1927)}

\section{Introduction}

A systematic treatment of environment, or natural resources, as a production factor is relatively recent in economic literature. The pioneering work of Harold Hotelling on 'the economics of exhaustible resources' \cite{Hotelling} had had a limited impact until the environmental awareness in the sixties, culminating with the publication of "Limits to Growth" (1972). Robert Solow's  much quoted Richard T. Ely Lecture \cite{Solow}, reflecting the current state of the art on the subject, was built on it. Solow's insights, in turn, have remained a reference in the mainstream literature, providing the theoretical arguments for the concept of `weak sustainability'.

The main issue he had proposed to deal with was not the exhaustibility of natural resources  in itself as a limit to the economic process, but the optimal social management of stocks of non-renewable but essential resources. The Laws of Thermodynamics were invoked at the onset to explain why materials recycling could not prevent the eventual exhaustion of all non-renewable resources and, indeed, why eventually the whole life on earth will come to an end.
Hotelling's rule was seen as a necessary condition for efficiency, and therefore for social optimality, but not a sufficient one. There were reasons for Solow to expect market interest rates to surpass the social rate of time preference, leading to an exceedingly fast market consumption of exhaustible resources. Therefore corrective public intervention should be brought in to slow down and stretch out exploitation of the resource pool. Moreover, there could also be several patterns of exploitation of the exhaustible-resource pool which obey Hotelling's fundamental principle myopically in the short term, but are wrong in the long term, implying markets need a reasonably accurate view of the long-term prospects.

Specifically, Solow pointed two conditions that could guarantee intergenerational equity in the access to the exhaustible-resource pool in the long run: first, the likelihood of technical progress, especially natural resource-saving advances - what is called today \emph{ecological efficiency} of the production process (e.g. some innovation reducing losses of ore in mining or smelting); second, the degree of substitutability of exhaustible resources by other factors of production, notably labor and reproducible capital (e.g. homeowners increasing expenditures on insulation to save on fuel costs, thus substituting fiberglass for heating oil). 

These are empirical issues, doubtless, but there would be grounds to expect prolonged and substantial reductions in natural-resource requirements per unit of real output as well as `quite a lot of substitutability between exhaustible resources and renewable or reproducible resources'. Solow stressed in particular the importance of the latter, specifically as to reproducible resources, reminiscent perhaps of the laws of thermodynamics he had evoked to point to the limits of the former. Since then, a huge amount of empirical research has been done to find evidence of them in the economic process `decoupling' from its natural resource basis. 

The ecological model behind Solow's analytical effort fails to address all relevant ecological aspects at stake: his Laws of Thermodynamics apply only to ecological efficiency; the continual substitution of new  natural resources for depleted ones has no thermodynamic consequence. Indeed that author neglects the second dimension of natural resources, as source of ecosystem services, which is inevitably impacted by the resulting increase of the mass-energy scale of the economic process.

What came to be known as `Economics of Pollution', as distinguished from `Economics of Exhaustible Resources', evolved to deal precisely with this second dimension. The work of Pigou on the concept of externalities \cite{Pigou} provided its theoretical basis. Economically, the specificity of this aspect derives from the absence of markets for ecosystem services, due to the public status of the ecosystems which provide them. To fix the problem, allowing markets to appear, the State must intervene either by  pricing or establishing property rights over  them. In so doing, market forces would lead to socially optimal paths of ecosystem services use, i.e. optimal pollution. There are no risks of potentially catastrophic losses of critical ecosystems in this model. 

Furthermore, the same conditions that could guarantee intergenerational equity in access to the exhaustible resources pool  in the long run would also imply equity of access to ecosystem services. Although renewable in principle, the latter can be treated as exhaustible because their stock at any given time is finite and can be destroyed by overuse. On the other hand, increases in ecological efficiency can prevent that, either reducing the waste generated by the productive process (`clean' technology), or substituting capital-intensive residue treatment facilities for the ecosystem service of waste absorption (`end-of-pipe' technology). Last, but indeed not least, even if an ecosystem is destroyed, regardless of its magnitude, its services could be replaced by equivalent ones produced by manmade devices. A rigorous description of how these conditions could allow for a (quasi-) perpetual economic growth was offered in \cite{Baumol}.

The lack of realism of the ecological model behind these analytical schemes was first made clear by Georgescu-Roegen in his `entropic' criticism of economic theory \cite{Georgescu}. As further elaborated by Daly and other ecological economists and ecologists\footnote{See \cite{Daly}, \cite{Costanza-Daly}, \cite{Daily}, \cite{Rockstrom2009a,Rockstrom2009b}, \cite{MEA} and \cite{Sukhdev}. See also \cite{Ozk-Ada-Dev} for a survey of the different ecological economics approaches. 
}, not only were the full implications of the Entropy Law to the economic process ignored, but in fact the very existence of \emph{critical ecosystem services}\footnote{To be fair, not all in the mainstream camp shared such unrealistic views on Nature. Bishop, based on the work of Ciriacy-Wantrup \cite{Ciriacy}, proposed a SMS (safe minimum standards) approach to handle the risk of irreversible losses of critical ecosystems \cite{Bishop}.}. Such services emanate from complex and interrelated ecosystems, being exceedingly difficult to replace by capital, and their depletion trajectories follow unforeseeable non-linear patterns due to the ecosystems' property of \emph{resilience}. The economic consequence of violating critical resilience thresholds is a sudden decrease in markets' own ability
to allocate mitigation efforts, thus opening the possibility of irreversible feedback mechanisms.

We derive in this paper  the theoretical implications of a realistic model for the natural impacts of the economic process. We incorporate elementary facts from Physics and Environmental Science into Baumol's orthodox perspective, which offers rigorous conditions for (quasi-)perpetual usage of resources and seems to allow  indefinite economic growth, even under the Entropy Law. Surprisingly, when considering Baumol's  model for resource depletion across all production processes in the economy, under the constraint of  critical ecosystem service stocks, we reach in fact the opposite conclusion. We establish the \emph{thermodynamic impossibility of indefinite growth in mass-energy scale of the economy} free from the risk of catastrophic environmental effects - an assumption commonly made, implicitly or explicitly, by most current growth models. We introduce instead the concept of \emph{responsible trajectory}, which offers a  quantitative criterion for sustainable economic policy.
%\newpage

In view of the inherent uncertainty around critical resilience thresholds, we prove that, even admitting the possibility of science fictional  substitutions of capital for natural assets, it is currently \emph{irresponsible} to enforce, or indeed to maintain, a trajectory of  material economic growth.  
\\

\noindent\textbf{Note on the use of Mathematics:} while our argument is quantitative, all claims are stated in plain English, keeping mathematical language to a universally accessible minimum and confining technical proofs only to Propositions \ref{prop def lower bound on w(t)} and \ref{prop P is bounded above}.
\\
\newline\noindent \quad\textbf{Acknowledgements:} Authors thank Dr. Joeri Rogelj from the Institute for Atmospheric and Climate Science at ETH Zurich for his comments on the draft version.

\newpage
\subsection{Initial definitions and statement of main results}

We begin with the following, hopefully uncontroversial, assumption:
\begin{axiom}   \label{axiom Universe}
Human economic activity takes place in the physical Universe, hence its process is constrained by the laws of Physics.
\end{axiom}
Fundamentally, the economic process transforms matter by means of mechanical, chemical and thermal devices which operate within a finite, albeit historically increasing, energy scale. It is therefore convenient to distinguish conceptually the subregion of the Universe where the economic process is a governing phenomenon: 

\begin{definition}
The \emph{anthroposphere} is the region of the Universe whose current state and dynamics are of the same order of magnitude as the economic process, hence subject to change under human design. \end{definition}

In other words, the anthroposphere is the total material substract of the economic process, together with the underlying interrelations of matter and energy, as imposed by Axiom \ref{axiom Universe}. However, two realms must be distinguished inside of the anthroposphere: the \emph{ecosphere}  where life exists and the \emph{abiotic sphere} exogenous to it. Thus the economic process can be understood as the mechanism by which matter-energy exchanges occur within the anthroposphere.
\begin{figure}[h]
%FIGURE 1
\[
\begin{xy}
,{\ar (10,30)*{}; (-10,30)*{}};  
,{\ar_{\txt{\tiny{mass-energy}}} (-10,29)*{}; (10,29)*{}};  
,(-30,20)*\txt{ecosphere}
,(30,20)*\txt{abiotic sphere}
,(0,15)*+\frm<150pt,80pt>{e}
,(0,42);(0,32)**@{--}
,(0,25);(0,9)**@{--}
,(0,1);(0,-12)**@{--}
,(0,5)*+={\txt{anthroposphere}}*\frm<80pt,30pt>{e}
,(0,-22)%*\txt{Physical Universe}
\end{xy}
\]
\caption{Physical Universe}
\end{figure}
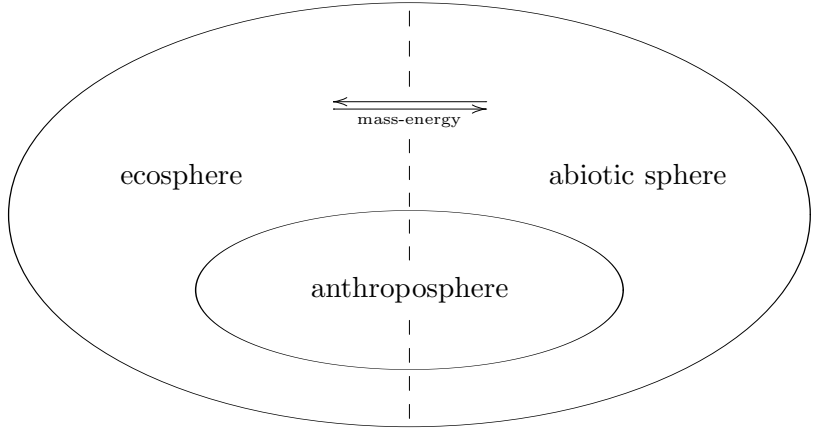

\begin{example}[boundaries of the anthroposphere and the ecosphere]
The Earth's atmosphere, land, biodiversity, fossil fuels and minerals, as well as atomic nuclei under interactions controlled by humans, are contained in the anthroposphere. Tidal, tectonic and planetary movements and stellar radiation are not, belonging strictly  in the abiotic sphere.

In particular, all the ecosystem services are contained in the ecosphere, and ecosystems are perturbed by the flow of mass-energy from the abiotic sphere. Such exchanges  may occur within or without the anthroposphere, i.e.,  may be respectively the result of human industry or natural phenomena (tectonics, meteorites etc). 
\end{example}

%\newpage
As an immediate consequence of Axiom \ref{axiom Universe}, the anthroposphere as a whole and the economic process  in particular are subject to thermodynamic constraints \cite{Kittel-Kroemer}, which we formulate as follows:
\begin{thermo}
Every transformation process based on mechanical, chemical or thermal devices, or combinations thereof, admits a definite upper bound on  efficiency, hence  a closed system of such processes has a net negative mass-energy balance with its surroundings, proportional to its scale.
\end{thermo}

We will give a precise meaning and proof to the following assertion:
\begin{mainthm} \label{thm growth contained by stock}
Assuming the Laws of Thermodynamics valid within a neighbourhood of the anthroposphere, one has:
\begin{enumerate}
        \item
        The growth in mass-energy scale of the economic process at time $t+1$ is constrained by the total stock of ecosystem services
at time $t$. 
        \item
        Any responsible growth trajectory, as assessed at time $t_0$, must predict,  within finite and possibly short time $t_1-t_0$, negative contributions to every type of high-entropy waste emission coming from low-entropy sources \emph{external} to the ecosphere. 
\end{enumerate}
\end{mainthm}

In face of the inherent uncertainty, or perhaps impossibility, in determining the actual critical resilience thresholds for all  essential ecosystem services, we arrive at a rather uncomfortable but logically necessary consequence:  
\begin{corzero}
Unless every critical  resilience threshold is accurately known for every ecosystem service indispensable to the economic process at a given time, the only responsible nonnegative growth trajectory is \emph{zero growth}.
\end{corzero}

%\newpage
\subsection{Baumol's solution to the depletion problem}
We adopt the definitions  from \cite{Baumol}, in mass-energy units, functions of a discrete time variable $t$ (measured e.g. in years):

%\vspace{-0.8cm}
\begin{center}
\begin{tabular}{cl}
$R_i(t)$ & usable stock of resource $i$ on Earth in period $t$.\\
$v_i(t)$ & quantity used up during period $t$. \\
$\displaystyle E_i(t)= \frac{1}{M_i(t)}R_i(t)$ & 
        \hspace{-10pt}
        \begin{tabular}{l}
        effective stock of resource $i$ in period $t$, where $M_i(t)>0$ is the \\ \emph{inefficiency ratio} of the production process w.r.t. resource $i$.         
        \end{tabular}
\end{tabular}
\end{center}
Note that recycling within the economy may allow $M_i$ to decrease below 1, i.e., some extracted resources are effectively consumed more than once. The extreme case $M_i(t)=0$ is the regime in which a resource $i$ can be reused indefinitely at zero loss, thus making the effective stock infinite regardless of the actual amount $R_i(t)$ still available.

Let $i^*$ denote the total number of resources involved in the economic process. In those terms, we formulate the relevant corollary of the Laws of Thermodynamics
which that author implements:

\begin{claim}[thermodynamic bound]      \label{claim thermo bound}
For every resource $1\leq i \leq i^*$, there is a definite lower bound $(M_i)_*>0$ for inefficiency, unsurmountable by innovation. 
\end{claim}
Since  $i^*$ is finite, the overall positive minimum $M_*=\min\limits_{1\leq i\leq i^*}(M_i)_*>0$ satisfies
\begin{equation}        \label{eq thermo lower bound M*}
M_i(t)\geq (M_i)_*\geq M_*>0, \qquad\forall i,t.
\end{equation}
%\vspace{-0.1cm}
Baumol proves that, even under the validity of Claim \ref{claim thermo bound}, every resource $i$ admits plausible depletion trajectories $\left(R_i(t),M_i(t)\right)$ such that the effective stock $E_i(t)$ remains constant or even increases over arbitrarily large (though ultimately finite) time intervals. It is thus established that the problem of growth under resource depletion reduces to maintaining suitable rates of innovation.

That author stops short, however, of considering the \emph{aggregate of such trajectories across the whole economy}, that is for all resources $1\leq i\leq i^*$, in the light of the same thermodynamic postulate understood at its full meaning. We take up the discussion from there, exploring the physically necessary effect of the production process on the environment's ability to provide services essential to the iteration of the process itself.
The irreversible nature of thermodynamic processes implies rather the opposite conclusion as to the long-term sustainability of the system, regardless of any innovations conceivable from the present stage of human inventiveness. Indeed, we will see that any sustainable trajectory, allowing the mere maintenance of our current production regime, let alone \emph{growth} in its mass-energy scale, is not only predicated on innovation but restricted by its exact marginal effective gain at any given time-period.
  
\newpage
\section{The economic process and the environment}
We wish to quantify the environmental impact of the economic process in terms of its irreversible deposits on the ecosphere, then incorporate  this data into Baumol's model. We define as \emph{waste} the matter-energy irremediably lost to entropy in the form of bound energy states, hence \emph{permanently excluded} from the production process, according to our  formulation of the Laws of Thermodynamics. Since recycled (or recyclable) materials are eventually reintroduced in the productive sphere, they are not considered waste for our purposes. The relevant conceptual framework is illustrated in the following scheme:
%\vspace{-1cm}
\begin{figure}[h]
%Figure 2
\[
\xymatrix{
\txt{Waste deposits \\(ecosphere)}&\fbox{\txt{Ecosystem stocks}}&&&& \\
\ar@{--}[rrrr] &&&&&\\
\txt{Production process \\(anthroposphere)} &\boxed{\txt{Natural sources}} \ar@/_2pc/[rr]_{production}% 
  && *[l]{\boxed{\txt{Economy}}} \ar@/_2pc/[ll]_{recycling}\ar@/_2pc/[uull]_{waste} \ar@(dr,ur)[]_{recycling}
&&\\
\ar@{--}[rrrr] &&&&&\quad\ar[uuu]|{\txt{Entropy}}\\
\\
}
\]
\vspace{-0.9cm}
\caption{Matter-energy in the production process}
\end{figure}
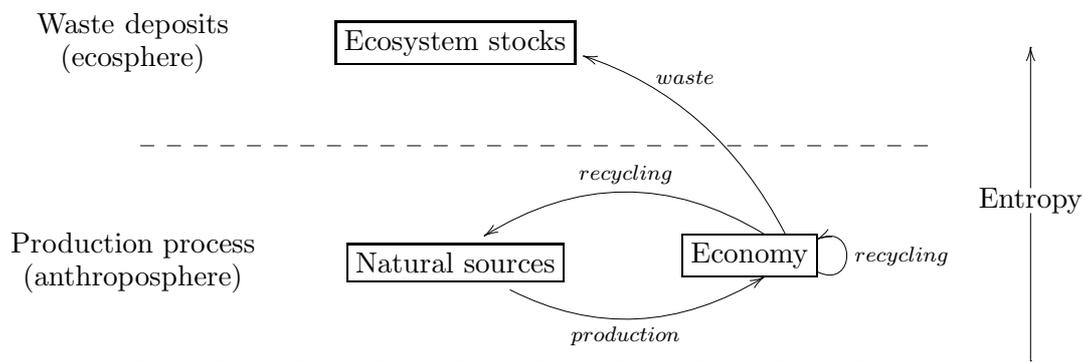

%\newpage
Let us begin with  the following notation:
\begin{center}
\begin{tabular}{rl}
$w_i(t)$ & waste generated from resource $i$ in period $t$.\\
$\phi(M)$& waste proportion at inefficiency ratio $M\geq0$.
\end{tabular}
\end{center}
The function  $\phi:\mathbb{R}^+\to [0,1]$ describes the proportion of an extracted resource which is wasted in the economic process at the given level of inefficiency. It is continuous, strictly increasing and satisfies $\phi(0)=0$ and $\displaystyle\lim_{M\to\infty}\phi(M)=1$, corresponding respectively to the extreme regimes of absolute efficiency and inefficiency. For our present purpose we do not need to know anything about $\phi$ beyond the previous properties, so let us adopt, without loss of generality, the simple model
$$
\phi(M):=\frac{M}{1+M}=1-\frac{1}{1+M}. 
$$
Then, in Baumol's notation for each resource $i$ and period $t$, we have:
\begin{equation}        \label{eq def waste w_i}
w_i(t)=\phi\left( M_i(t)\right)\cdot v_i(t).
\end{equation}

\newpage
\subsection{Stock of ecosystem services}

\begin{definition}
A \emph{(critical) ecosystem service} is the ecosphere's capacity (in mass-energy units) to absorb bound states of matter of a given chemical type, or aggregates thereof,  without impairing significantly the current economic process.
\end{definition}

We need therefore the following additional definitions:
\begin{center}
\begin{tabular}{rl}
$\sigma_j(t)$ & remaining usable stock of ecosystem service $j$ at time $t$.\\
$w_{ij}(t)$& amount of waste from resource $i$ that contributes to degrade service $j$.\\
$W_j(t)$& total amount of service $j$ degraded during period $t$.\\
$j^*$& number of known ecosystem services.
\end{tabular}
\end{center}

Hence, at time  $t$, we denote the total waste generated from resource $i$ 
$$
w_{i\sharp}(t):=\sum_{j=1}^{j^*} w_{ij}(t)
$$
and the total degradation of ecosystem stock $j$, \emph{due to newly extracted resources}, 
\begin{eqnarray}        \label{eq total degradation of stock j}
w_{\sharp j}(t):= \sum_{i=1}^{i^*} w_{ij}(t).
\end{eqnarray}
The fact that not all ecosystem degradation comes from newly extracted resources is expressed by 
$$
W_j(t)\geq \sum_{j=1}^{j^*}  w_{\sharp j}(t),
$$
hence, tautologically, 
$\sigma_j(t+1)=\sigma_j(t)-W_j(t)\leq\sigma_j(t)-w_{\sharp j}(t).$

%\newpage
\subsection{Physical bound on effective ecosystem services}
Inefficiency has a lower bound, hence proportional waste is inevitable. We will see that, in such terms, accumulated waste eventually exceeds the finite stock of ecosystem services.
\begin{proposition}     \label{prop def lower bound on w(t)}
To every increase in the mass-energy scale of the economic process at time $t$, there corresponds an inevitable reduction of some ecosystem stocks, proportional in total to that increase at least, occurring at time $t+1$.
\begin{proof}
In the terms of Claim \ref{claim thermo bound} and   (\ref{eq def waste w_i}), write $w_*:=\phi\left(M_*\right)>0$ for the waste proportion at optimum efficiency, as well as  
\begin{equation}        \label{eq def v(t) e w(t)}
v(t):=\sum\limits_{i=1}^{i^*}v_i(t)\quad \text{and} \quad w(t):=\sum\limits_{i=1}^{i^*}w_{i\sharp}(t).
\end{equation}
One obtains trivially from (\ref{eq thermo lower bound M*})  that $
w_{i\sharp}(t)\geq w_*v_i(t),\; \forall i,t.
$
Denoting
$\displaystyle
W(t) :=\sum_{j=1}^{j^*} W_j  (t)
$ 
the \emph{total waste} produced at time $t$,
i.e., the total reduction of ecosystem stocks at time $t$, we find
\begin{equation*}
W(t)\geq w(t)\geq w_*v(t),\quad \forall t.\qedhere
\end{equation*}
\end{proof}
\end{proposition}
Thus the amount of entropic waste corresponding to every tentative increase in the total mass-energy is not only strictly positive, but indeed at least proportional to the amount of matter-energy entered into the cycle.

Define the  \emph{mass-energy scale} of the economic process at time $t$:
\begin{equation}        \label{eq def P(t)}
P(t):=\sum\limits_{\hat t=-\infty}^{t}\left(v(\hat t)-w(\hat t)\right).
\end{equation}

%\newpage
\begin{proposition}     \label{prop mass contribution}
In the absence of external sources of low entropy, there exists a time-independent constant $q_*>0$ such that
$$
w(t+1)\geq w_*v(t+1)+q_*P(t).
$$
\begin{proof}
The total mass-energy circulating in the economic process is transformed, by assumption, under mechanical, chemical and thermal interactions, all of which have definite upper-bounds on efficiency, by our physical assumptions. The extensivity property of the entropy function \cite{Lieb-Yngvason1998, Lieb-Yngvason1999} implies that the aggregate of all such processes, considered as a closed system, can operate no more effectively than the most efficient of its constituent processes, hence a definite proportion of the mass-energy in the economic process dissipates unavoidably into the ecosphere, in the form of bound energy states. 
\end{proof}
\end{proposition}

To make that crucial point crystal-clear, if the total amount of matter-energy  contained the economic process at time $t$ were still entirely contained in the system at time $\left.t+1\right.$,   then \emph{the whole economic process would be a thermodynamical system in perpetual motion}, thus violating the physical constraint. 
 
%\newpage
\subsection{Insufficiency of markets under ecological thresholds}
\label{subsec thresholds}
One might argue at this stage that markets' resourcefulness in face of impending exhaustion of an ecosystem stock would suffice to prevent the actual crossing of the resilience threshold, e.g. by pricing mechanisms which penalise offenders and generate enough incentive towards new alternatives. Unfortunately this objection relies on the fallacy of preemptive knowledge of such thresholds in complex systems. 

The fundamental notion here is the epistemological phenomenon of \emph{novelty by combination} \cite{Georgescu}, the fact that composite systems develop qualitative characteristics which are \emph{a priori inaccessible} even from complete knowledge of its constituent subsystems. For example, it is simply not possible to derive the point at which water transitions from its liquid to gaseous forms by any calculation, using as input some chemical data of the oxygen and hydrogen atoms alone, which eventually returns $100\textordmasculine\ C$. The boiling point of water is indeed an empirical fact, determined through repeated experimental observation.   

Therefore, it is entirely possible that the actual value of an ecosystem stock is \emph{beyond assessment} by predictive deduction, and can only be quantified empirically, i.e. \emph{once it is violated}.  This means any pricing solution to the use of the remaining stock of such a resource will be essentially  anyone's guess. 

Furthermore, such phase transitions might lead to irreversible changes in the environment, towards a new equilibrium state which impairs markets' very ability to supply equivalent ecosystem services by substituting for, say, additional labour or capital. 
\begin{example}[definite ecosystem service stock]
While ozone gas concentration at the high atmosphere provides useful shielding against stellar radiation, the same ozone at the lower atmosphere interacts strongly with other molecules and is extremely toxic to living systems. The total stock of ozone absorption at the lower atmosphere is a known  quantity, beyond which such chemical interactions harm crop production, forest growth  and human health \cite{WMO}.

In the event that our emissions cross the boundary and effects gradually incur, it is indeed possible that the new demand will generate technological and managerial innovations to counter them. However, (i) when taking for granted that
deterioration will maintain  gradual pace one overlooks  the potential
feedback effect that our unregulated mitigation efforts might have on further
emissions and aggravation of the phenomenon; and in any case, crucially, (ii) it
remains  to consider whether significant progress can be reached at a \emph{higher
rate} than e.g. the increased healthcare costs and agricultural losses degrade
our productive capacities.

\end{example}
The above example, which is far from exceptional, illustrates how the exhaustion of an ecosystem stock, at time $t$, can henceforth hinder the economic process' normal allocation functions, as assessed  at time $t-1$. If, moreover, one brings into account the non-linear trajectories of ecosystem depletion, then it is quite possible that, at an unknown point thereafter, the protective ecosystem will suddenly collapse, with catastrophic effects to the economic process, as understood at time $t-1$.

\newpage
\section{Low-entropy trajectories versus irresponsible growth}
We formulate a precise quantitative meaning for the responsibility of economic trajectories and prove  the
\textbf{Theorem of limited growth}. In these terms, any responsible growth in the material scale of the economy must be authorised by actual achievements in innovation which increase the effective stock of ecosystem services. Bringing in the uncertainty of resilience thresholds, we deduce the \textbf{Corollary of zero growth}. 
 
\subsection{Responsible and irresponsible trajectories}

Recall from (\ref{eq total degradation of stock j}) that $w_{\sharp j}(t)$ denotes the total degradation of ecosystem service $j$, due to new material sources, at time $t$. We now relate qualitatively the sum of such contributions over time to the constraint posed by finite ecosystem stocks.

\begin{definition}      \label{def responsible trajectory}
A time-trajectory of the economic process will be deemed \emph{responsible} at time $t_0$ if
$$
\sum\limits_{t\geq t_0}w_{\sharp j}(t)\leq \sigma_j(t_0), \quad \forall 1\leq j\leq j^*. 
$$ 
A trajectory is \emph{irresponsible} at $t_0$ if it cannot be deemed responsible. \end{definition}

A responsible trajectory degrades every ecosystem service $j$, at every  $t\geq t_0$, by an amount $w_j(t)$ such that the total future waste converges to a value no greater than the total stock $\sigma_j(t_0)$ of service $j$ at $t_0$.
In any other case the trajectory is irresponsible; be it because it is known to violate some threshold $\sigma_j$, or indeed \emph{because that is not known} for sure. This semantics is in strict colloquial accordance, in the sense that a morally accountable attitude   presupposes reasonable certitude of its consequences. For instance, it is irresponsible to hire a babysitter known to be untrustworthy; but the same would be said of hiring a complete stranger, whose degree of trust cannot be ascertained at present. Similarly, a responsible growth strategy must foresee the perpetuation of the economic process itself before its ultimately moral purpose of promoting the welfare of Humanity, hence it cannot allow even the \emph{risk} of irreversible systemic collapse.  

\begin{proposition}     \label{prop P is bounded above}
Along any responsible trajectory, $P(t)$ is bounded above.

\begin{proof}
Taking the infinitesimal form of the mass-energy scale (\ref{eq def P(t)}) we have:
$$
P'(t)=v(t)-w(t)\leq\left( \frac{1}{w_*} -1\right)w(t),\quad\forall t.
$$
Set, for clarity, $\displaystyle C:=\frac{1}{w_*}-1>0$. Integrating over any interval $[t_0,t]$, we get 
$$
P(t)-P(t_0)  \leq  C\int_{[t_0,t]} w \leq  C\int_{[t_0,\infty]} w.
$$
If, by contradiction, there exists a sequence $\left(t_n\right)\subset[t_0,\infty]$ such that $\lim\limits_{n\to\infty} P(t_n)=\infty$, then the integral on the r.h.s. must diverge, hence by the definition of $w(t)$ [see (\ref{eq def v(t) e w(t)})] there is at least one pair of  indices  $(i,j)$ such that $$
\int_{[t_0,\infty]} w_{ij}\to \infty.
$$
Therefore $\sum\limits_{t\geq t_0}w_{\sharp j}(t)$ diverges and the trajectory one considers is irresponsible. 
\end{proof}
\end{proposition}
\noindent
Together with Proposition \ref{prop def lower bound on w(t)}, this proves the \textbf{Theorem of limited growth}. 

Finally, since the condition for responsibility [see \emph{Definition \ref{def responsible trajectory}}] requires the knowledge at time $t$ of all the quantities $\sigma_j (t)$, we must infer that every positive growth trajectory is necessarily irresponsible if so much as one of those thresholds remains undetermined; this is the \textbf{Corollary of zero growth}. In  the light of \emph{Subsection \ref{subsec thresholds}}, we are led to assert that novelty by recombination  implies (at best) zero growth in the matter-energy scale of the economy, at any time $t$. 

%\newpage
\subsection{Conclusion}

Although there is no physically meaningful way to avoid the ultimate consequences of Proposition \ref{prop P is bounded above}, a few back doors can be proposed. First, a rather pedantic but logically sound point is the fact that the economy can indeed grow perpetually in ways which do not imply a significant increase in matter-energy scale, e.g. by the production of knowledge, culture and art. In any case, this excludes long-term consumerist ambitions, and such path would likely imply an even more revolutionary change in the way we understand Economics than the alternatives to follow.  

Second, and most important, it would be possible to raise the upper bound on the mass-energy scale if the economic process could access sources of low entropy \emph{exterior} to the ecosphere, hence genuinely \emph{recycle} bound energy states from the environment (or \emph{dispose} of them very far away) and restore the net stocks of ecosystem services. Mathematically this would correspond to a negative term on the right-hand side in Proposition \ref{prop mass contribution}, thus in principle  allowing responsible trajectories with indefinitely increasing effective mass-energy scales. 

Such technological solutions, however, remain in the realm of science fiction.  If engineers of the future do make the necessary advances, in the form of low-entropy sinks operational at the magnitudes of global production, then the authors of this paper - or rather our descendants - will gladly see our conclusions obsolete, in the eyes of the economists and the public of that future. Meanwhile, to leverage present consumption on the prospective development of futuristic solutions is morally tantamount to a North-American economist, say, gambling their family's education savings in the derivatives market - which hardly any one would do, regardless of their ideological inclinations. 

In face of our current irresponsible trajectory, Thermodynamics raises  a persuasive warning. Foreseeable outcomes of  a collapse of the economic process include worldwide destitution and conflict; the very scenarios envisioned by those who see in perpetual growth the only way to keep social peace.

%\newpage
\bibliographystyle{agsm}
\bibliography{Entropy_and_Growth}

\end{document}